\documentclass[10pt,twocolumn,letterpaper]{article}

\usepackage[pagenumbers]{cvpr} 

\usepackage{booktabs}
\usepackage{amsmath}
\usepackage{amssymb}
\usepackage{bm}
\usepackage{url}
\usepackage{ragged2e}
\usepackage{multirow}
\usepackage{tablefootnote}
\usepackage{xspace}
\usepackage{xcolor,colortbl}
\usepackage{wrapfig}
\usepackage{makecell}
\usepackage[thicklines]{cancel}
\usepackage{xcolor}

\usepackage{tcolorbox}
\usepackage{bbding}
\usepackage{pifont}
\usepackage{arydshln} 
\usepackage{lipsum}
\usepackage{float}
\usepackage{booktabs}
\usepackage{utfsym}
\usepackage{subcaption}

\makeatletter

\def\@IEEEsectpunct{.\ \,}
\def\paragraph{\@startsection{paragraph}{4}{\z@}{1.0ex plus 1.ex minus 0.5ex}%
{0ex}{\bfseries}}%
\makeatother

\definecolor{lightgray}{HTML}{F5F5F5} 
\usepackage[table]{xcolor} 
\definecolor{lightblue}{RGB}{210,222,245}  
\definecolor{lighterblue}{RGB}{232,241,252} 

\newcommand{\highlightblue}[1]{\colorbox{lightblue}{#1}}
\newcommand{\highlightlightblue}[1]{\colorbox{lighterblue}{#1}}
\newcommand{\methodname}{HiCoDiT\xspace}

\usepackage{enumitem}

\definecolor{maroon1}{RGB}{0,0,0} 

\usepackage{bm}
\DeclareMathAlphabet{\mathsfit}{\encodingdefault}{\sfdefault}{m}{sl}
\SetMathAlphabet{\mathsfit}{bold}{\encodingdefault}{\sfdefault}{bx}{n}

\def\mP{{\bm{P}}}
\def\mQ{{\bm{Q}}}

\makeatletter

\DeclareRobustCommand{\cev}[1]{%
  {\mathpalette\do@cev{#1}}%
}
\newcommand{\do@cev}[2]{%
  \vbox{\offinterlineskip
    \sbox\z@{$\m@th#1 x$}%
    \ialign{##\cr
      \hidewidth\reflectbox{$\m@th#1\vec{}\mkern4mu$}\hidewidth\cr
      \noalign{\kern-\ht\z@}
      $\m@th#1#2$\cr
    }%
  }%
}

\definecolor{cvprblue}{rgb}{0.21,0.49,0.74}
\usepackage[pagebackref,breaklinks,colorlinks,allcolors=cvprblue]{hyperref}

\def\confName{CVPR}
\def\confYear{2026}

\title{Hierarchical Codec Diffusion for Video-to-Speech Generation}

\author{Jiaxin Ye$^{1}$, Gaoxiang Cong$^{2,3}$, Chenhui Wang$^1$, \\Xin-Cheng Wen$^4$, Zhaoyang Li$^1$, Boyuan Cao$^1$, Hongming Shan$^{1}$\footnotemark[2] \\
$^1$Fudan University, 
$^2$Institute of Computing Technology, Chinese Academy of Sciences\\
$^3$University of Chinese Academy of Sciences,
$^4$Harbin Institute of Technology (Shenzhen)\\
{\tt\small jxye22@m.fudan.edu.cn, hmshan@fudan.edu.cn}
}

\renewcommand{\paragraph}[1]{\noindent\textbf{#1}\quad}

\begin{document}
\maketitle

\renewcommand{\thefootnote}{\fnsymbol{footnote}}
\footnotetext[2]{Corresponding author.}
\renewcommand{\thefootnote}

\begin{abstract}

Video-to-Speech (VTS) generation aims to synthesize speech from a silent video without auditory signals. 
However, existing VTS methods disregard the hierarchical nature of speech, which spans coarse speaker-aware semantics to fine-grained prosodic details. This oversight hinders direct alignment between visual and speech features at specific hierarchical levels during property matching.
In this paper, leveraging the hierarchical structure of Residual Vector Quantization (RVQ)-based codec, we propose \textbf{\methodname}, a novel \textbf{Hi}erarchical \textbf{Co}dec \textbf{Di}ffusion \textbf{T}ransformer that exploits the inherent hierarchy of discrete speech tokens to achieve strong audio-visual alignment.
Specifically, since lower-level tokens encode coarse speaker-aware semantics and higher-level tokens capture fine-grained prosody, \methodname employs low-level and high-level blocks to generate tokens at different levels. The low-level blocks condition on lip-synchronized motion and facial identity to capture speaker-aware content, while the high-level blocks use facial expression to modulate prosodic dynamics.
Finally, to enable more effective coarse-to-fine conditioning, we propose a dual-scale adaptive instance layer normalization that jointly captures global vocal style through channel-wise normalization and local prosody dynamics through temporal-wise normalization.
Extensive experiments demonstrate that \methodname outperforms baselines in fidelity and expressiveness, highlighting the potential of discrete modelling for VTS. 
The code and speech demo are both available at {\href{https://github.com/Jiaxin-Ye/HiCoDiT}{https://github.com/Jiaxin-Ye/HiCoDiT}}.

\end{abstract}

\section{Introduction}
\label{sec:intro}

Video-to-Speech (VTS)~\cite{demoface/abs-2502-01046,visualvoicecloning/Cong0QZWWJ0H23,DBLP:conf/mm/00010PZBHQH25} generation aims to infer and synthesize speech from visual cues alone. This capability enables transformative applications, including silent film dubbing and assistive communication for aphonic individuals, to seamless interaction in noise-sensitive~\cite{darur2025visual}, privacy-critical~\cite{potamianos2004audio} or embodied~\cite{feng2026self} environments. 

The fundamental challenge in VTS lies in addressing the inherent information asymmetry between visual and acoustic modalities when generating natural and lip-synchronized speech from visual without acoustic input guidance. Specifically, although facial video and speech share consistent content~\cite{whisper/RadfordKXBMS23}, identity~\cite{speechbrain}, and emotional prosody~\cite{timnet:conf/icassp/YeWWXLS23,emodna:conf/mm/YeWWMH0S23,ctlmtnet:conf/ijcai/WenYLXWW022}, visual features are inherently sparse and insufficient to capture the dense representations of speech, making it difficult to build accurate cross-modal alignment. 

Existing approaches predominantly focus on representation alignment for guiding generative models~\cite{diffv2s:conf/iccv/ChoiHR23,Face2Speech:conf/interspeech/GotoOSTM20,facespeak:conf/cvpr/JangKAKYJKKC24,visualvoicecloning/ChenTQZLW22}, spanning semantic content, vocal identity, and emotional prosody from vision to speech: (\emph{i}) for semantic content alignment, NaturalL2S~\cite{naturall2s/nn/LiangLLLLZ26} leverages multimodal self-supervised representations to enhance the alignment between visual semantics and speech content; (\emph{ii}) for vocal identity alignment, Face2Speech~\cite{Face2Speech:conf/interspeech/GotoOSTM20} aligns features from a face recognition encoder and a speaker recognition encoder to map facial identity to timbre information, while Face-StyleSpeech~\cite{facestylespeech:journals/corr/abs-2311-05844} further incorporates contrastive learning to improve face-to-speech alignment; (\emph{iii}) for emotional prosody alignment, FTV~\cite{FTV:conf/cvpr/KimCKJC25} aligns facial emotion embeddings with pitch and energy to enhance prosody expressiveness. 
However, existing VTS methods typically inject visual features into holistic speech representations while overlooking the hierarchical structure of speech, from coarse speaker-aware semantics to fine-grained prosodic details, which ultimately exacerbates the inherent information asymmetry between visual and acoustic modalities. 
Therefore, the principal bottleneck in high-quality video-to-speech generation lies in visual conditioning, and how to exploit the speech hierarchy as a prior to improve generation quality remains unresolved.

In this paper, we propose \textbf{\methodname}, a novel \textbf{Hi}erarchical \textbf{Co}dec \textbf{Di}ffusion \textbf{T}ransformer that fully leverages the inherent hierarchy of discrete speech tokens to enable more effective vision-speech alignment. 
To the best of our knowledge, \methodname is the first to introduce an explicit speech hierarchy prior into a discrete diffusion framework for video-to-speech generation.
Specifically, leveraging the hierarchical structure of the Residual Vector Quantization (RVQ) codec in Figure~\ref{fig:rvq}, the low-level tokens primarily capture rich speaker-aware semantic content, whereas the high-level tokens encode more abstract prosodic details. Therefore, the hierarchy prior dictates that visual features such as lip motion and facial identity should primarily refine low-level speech tokens, while facial emotion features should modulate high-level tokens. 
Motivated by this prior, we design a hierarchical codec diffusion transformer composed of low-level and high-level blocks, progressively conditioning on speech tokens across different levels. 
The low-level blocks generate tokens conditioned on synchronized lip-motion representations and facial identity features for semantic and timbre alignment, while the high-level blocks produce tokens guided by facial emotion sequences for prosody alignment.
To achieve more effective conditioning in the high-level block, we introduce a dual-scale Adaptive Instance Layer Normalization (AdaLN) that employs channel-wise normalization to model global vocal style, and temporal-wise normalization to capture local prosody dynamics. 
Extensive experiments demonstrate that \methodname surpasses state-of-the-art baselines in semantic alignment and expressive prosody. 
Our contributions are summarized as follows.
\begin{itemize}[leftmargin=12pt]
\item To our knowledge, \methodname is the first discrete diffusion framework for VTS to explicitly integrate speech hierarchy prior, bridging the gap between video and speech. 
\item We propose a novel hierarchical diffusion transformer that models the speech hierarchy while disentangling visual conditioning, and a  dual-scale AdaLN to inject global vocal style and local prosody into speech generation, enhancing expressiveness and fidelity. 
\item Extensive experiments demonstrate superior performance in semantic consistency and speech diversity, highlighting the potential of discrete speech tokens modelling for efficient VTS generation.
\end{itemize}

\section{Related Work}
\label{sec:relatedwork}

\paragraph{Video-to-Speech (VTS) generation.}
Video-to-speech (VTS) seeks to generate speech that accurately reflects both linguistic content and speaker identity from visual cues alone~\cite{diffv2s:conf/iccv/ChoiHR23,lipvoicer:conf/iclr/YeminiSBGF24,MTL_lip:conf/icassp/KimHR23}. Current approaches typically enforce alignment through auxiliary objectives: some predict text or mel-spectrograms jointly with visual input~\cite{MTL_lip:conf/icassp/KimHR23}, others condition speaker embeddings on lip motion~\cite{diffv2s:conf/iccv/ChoiHR23} or minimize cross-modal embedding distances~\cite{Face2Speech:conf/interspeech/GotoOSTM20}. While effective in isolation, these methods treat speech as a flat sequence without hierarchy and impose multiple supervision signals, leading to suboptimal alignment. 

Recent advances have explored more sophisticated generative frameworks. For example, FTV~\cite{FTV:conf/cvpr/KimCKJC25} employs flow matching with a hierarchical visual encoder to gradually inject visual features into continuous mel-spectrogram space, while VoiceCraft-Dub~\cite{voicecraft:journals/corr/abs-2504-02386} adapts pretrained autoregressive discrete text-to-speech models~\cite{VoiceCraft:conf/acl/Peng00MH24} to incorporate visual context. However, both obscure the inherent hierarchical structure of speech representation, in which coarse linguistic content emerges at early token levels and fine prosodic detail is resolved later. In contrast, \methodname introduces the first discrete diffusion model for VTS trained from scratch, which explicitly integrates the speech hierarchy prior, bridging the gap between video and speech.

\paragraph{Hierarchical speech generation.}
Given the intrinsic hierarchy of speech, extensive research has focused on hierarchical representation modelling to achieve high-quality speech generation. 
In text-to-speech (TTS), Lee~\etal~\cite{hierspeech:conf/nips/LeeKLSHL22} propose a hierarchical conditional variational autoencoder (VAE) that leverages self-supervised speech representations to bridge the information gap between text and speech, and Hsu~\etal~\cite{HGMTTS:conf/iclr/HsuZWZWWCJCSNP19} likewise propose a conditional VAE with two levels of hierarchical latent variables, which captures coarse acoustic information and refines specific attribute configurations. 
Similarly, in video-to-speech, Kim~\etal~\cite{FTV:conf/cvpr/KimCKJC25} develop a hierarchical visual encoder that learns a conditional representation by progressively aligning content, timbre, and prosody for a flow-matching decoder. 
In contrast to prior works that design entangled conditioning for hierarchical speech attributes, we exploit the inherent hierarchy of speech tokens themselves, modelling speech tokens from coarse semantics at lower levels to fine-grained acoustic details at higher levels, enabling disentangled conditioning and improving the fidelity of speech generation.

\section{Preliminary: Discrete Diffusion Models}
\label{subsec:discrete_diffusion}

Recently, continuous diffusion models (CDM)~\cite{mixganTTS:journals/access/DengWQLC23,diffv2s:conf/iccv/ChoiHR23,DBLP:journals/ijautcomp/LuZBCLZ25,DBLP:journals/ijautcomp/ZhangWHWWZZLL25,wang2026brain} have achieved state-of-the-art results in multimedia generation~\cite{zhou2026spatialrewardverifiablespatialreward,zhou2026unifiedthinkergeneralreasoning,caorepldm,DBLP:journals/corr/abs-2410-13830,DBLP:conf/ijcai/WangCCHJWS24}, while they are limited by computational inefficiency, frustrating practical application. An intuitive solution is to utilize discrete speech tokens~\cite{encodec:journals/tmlr/DefossezCSA23,Srcodec:conf/icassp/ZhengTXX24,maskgct:journals/corr/abs-2409-00750} to build discrete diffusion models (DDMs), which have shown potential in language modeling~\cite{D3PM:conf/nips/AustinJHTB21,ConcreteScoreMatch:conf/nips/MengCSE22,SEDD:conf/icml/LouME24} and speech generation~\cite{diffsound:journals/taslp/YangYWWWZY23,DCTTS:conf/icassp/WuLLY24}. 
In this paper, we introduce a masked-based DDM to generate speech tokens under cross-modal guidance and outline below the forward and reverse processes of the DDM, along with its training objective.

\paragraph{Forward diffusion process.}
Given a token sequence $\bm{x} = [x^1, \ldots, x^d]$ with length $d$, where each token belongs to a discrete state sapce $\mathcal{X} = \{1, \ldots, n\}$. The diffusion process can be modelled as a continuous-time discrete Markov chain, parameterized by the diffusion matrix $\mQ_t\in \mathbb{R}^{n^d\times n^d}$, also known as the transition rate matrix at time $t$, as follows:
\begin{equation}
\label{eq:delta_transition}
    p({x}_{t+ \Delta t}^i|x_t^i) =  \delta_{{x}^i_{t+ \Delta t}x^i_t} + \mQ_t({x}^i_{t+ \Delta t},x^i_t)\Delta t + o(\Delta t),
\end{equation}
where $\delta$ is Kronecker delta, $x^i_t$ denotes $i$-th element of $\bm{x}_t$, and $\mQ_t({x}^i_{t+ \Delta t},x^i_t)$ is the $({x}^i_{t+ \Delta t},x^i_t)$ element of $\mQ_t$, which represents the transition rate from state $x^i_t$ to state ${x}^i_{t+ \Delta t}$ at time $t$. To further achieve efficient computation, existing methods~\cite{SEDD:conf/icml/LouME24,RADD:journals/corr/abs-2406-03736} adopt the assumption of dimensional independence, conducting a one-dimensional diffusion process for each dimension with the same token-level diffusion matrix $\mQ_t^\text{tok}=\sigma(t)\mQ^\text{tok}\in \mathbb{R}^{n\times n}$, where $\sigma(t)$ is the noise schedule and $\mQ^\text{tok}$ is designed to diffuse towards an masked state \texttt{[MASK]}. 
Now, the forward equation can be formulated as $\mP({x}^i_t,x^i_0) = \exp\left(\bar{\sigma}(t) \mQ^\text{tok}({x}^i_t,x^i_0) \right)$, where transition probability matrix $\mP({x}^i_t,x^i_0) := p({x}^i_t|x_0)$, and cumulative noise $\bar{\sigma}(t) = \int_0^t \sigma(s)ds$. There are two probabilities in the $\mP_{t|0}$: \( 1 - e^{-\bar{\sigma}(t)} \) for replacing the current tokens with \texttt{[MASK]}, \( e^{-\bar{\sigma}(t)} \) for remaining unchanged. 
Finally, the corrupted sequence $\bm{x}_t$ can be sampled from $\bm{x}_0$ in one step.

\paragraph{Reverse unmasking process.}
Given the diffusion matrix $\mQ^\text{tok}_t$, we need a reverse transition rate matrix $\bar{\mQ}_t$~\cite{SCDDM:conf/iclr/SunYDSD23,kelly2011reversibility} to formulate reverse process, where $\bar{\mQ}_t(x^i_{t- \Delta t},x^i_t)=\tfrac{ p(x^i_{t- \Delta t})}{p(x^i_t)} \mQ^\text{tok}_t(x^i_t,x^i_{t- \Delta t})$ and $x^i_{t- \Delta t}\neq x^i_t$, or $\bar{\mQ}_t(x^i_{t- \Delta t}, x^i_t) =  - \sum_{z \neq x_t} \bar{\mQ}_t(z,x^i_t)$. 
The reverse equation is formulated as follows: 
\begin{equation}
    \label{eq:backward}
    p({x}^i_{t- \Delta t}|x^i_t) =  \delta_{{x}^i_{t- \Delta t}x^i_t} + \bar{\mQ}_t({x}^i_{t- \Delta t},x^i_t)\Delta t + o(\Delta t).\\
\end{equation}
The core of the reverse unmasking process is to estimate the concrete score $c_{x^i_{t- \Delta t} {x}^i_t}=\tfrac{p(x^i_{t- \Delta t})}{p(x^i_t)}$ of $\bar{\mQ}_t$, representing to measure the \textit{transition probability or closeness} from a state $x^i$ at time $t$  to a state $\hat{x}^i$ at time $t- \Delta t$. We can introduce a score network $s_\theta({x}^i_t,t)_{x^i_{t- \Delta t}} \approx [\tfrac{p(x^i_{t- \Delta t})}{p(x^i_t)}]_{x^i_{t}\neq x^i_{t- \Delta t}}$ to learn the score, so that the reverse matrix is parameterized to model the reverse process $q_\theta({x}^i_{t- \Delta t}|x^i_t)$ (\ie,~parameterize the concrete score).

\paragraph{Training objective.}
Denoising score entropy (DSE)~\cite{SEDD:conf/icml/LouME24} is introduced to train the score network $s_\theta$:
\begin{equation}
\label{eq:score_entropy}
\begin{aligned}
       \int_0^T \mathbb{E}_{\bm{x}_t \sim p\left(\bm{x}_t \mid \bm{x}_0\right)} \sum_{{\hat{\bm{x}}_t} \neq \bm{x}_t} \mQ_t\left( \hat{{x}}^i_t,{x}^i_t\right)  \Big[s_\theta\left({x}^i_t, t\right)_{\hat{{x}}^i_t}  & \\
        - c_{\hat{{x}}^i_t {x}^i_t} \log s_\theta\left({x}^i_t, t\right)_{\hat{{x}}^i_t}+  \text{N}(c_{\hat{{x}}^i_t {x}^i_t})\Big] dt &,
\end{aligned}
\end{equation}
where the concrete score $c_{\hat{{x}}^i_t {x}^i_t} = \tfrac{p\left({\hat{{x}}^i_t} \mid {x}^i_0\right)}{p\left({x}^i_t \mid {x}^i_0\right)}$ and a normalizing constant function $\text{N}(c):= c \log c - c$ that ensures loss non-negative. 
During sampling, we start from $\bm{x}_{T}$ filled with masked token \texttt{[MASK]}, and iteratively sample new set of tokens $\bm{x}_{t-1}$ from $p_{\theta}(\bm{x}_{t-1}|\bm{x}_{t})$ by replacing the concrete score with the trained score network on~\cref{eq:backward}.

\section{Methodology}
\label{sec:method}

\begin{figure*}[t]
    \centering
    \includegraphics[width=0.99\linewidth]{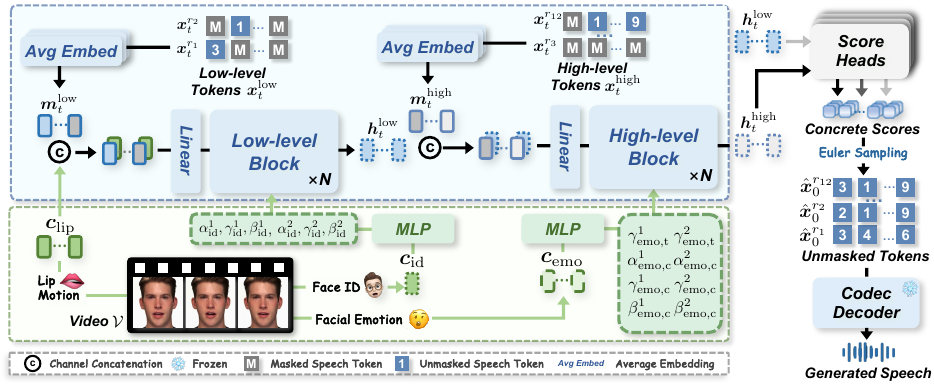}
    \caption{\textbf{Overall framework of \methodname}. We formulate video-to-speech generation as a hierarchical masked token prediction task. Speech is tokenized using an RVQ codec and split into low-level components $\bm{x}^{r_1:r_2}_t$ and high-level components $\bm{x}^{r_3:r_{12}}_t$, reflecting the intrinsic hierarchy of speech tokens. Guided by this structure, we disentangle visual features from the input video $\mathcal{V}$ into lip motion $\bm{c}_\text{lip}$, identity $\bm{c}_\text{id}$, and emotion $\bm{c}_\text{emo}$, and inject them into the corresponding diffusion blocks. Finally, score heads take output features $\bm{h}_t^{\text{low}}$ and $\bm{h}_t^{\text{high}}$ from both blocks to predict concrete scores of all level tokens for unmasking. }
    \label{fig:main}
\end{figure*}

\subsection{Overview of \methodname}
\label{subsec:overview}

Given a silent video $\mathcal{V}$, the goal of VTS system is to synthesize high-fidelity speech that aligns with the extracted visual features from the input video, including lip motion $\bm{c}_\text{lip}$, identity $\bm{c}_\text{id}$, and emotional expression $\bm{c}_\text{emo}$. 
To explicitly integrate speech hierarchy prior, we formulate VTS as a hierarchical masked token prediction task, employing an RVQ codec to tokenize speech for high-fidelity generation~\cite{encodec:journals/tmlr/DefossezCSA23} and a discrete diffusion model to decode masked tokens for strong in-context perception~\cite{SEDD:conf/icml/LouME24}. 
Specifically, as shown in Figure.~\ref{fig:main}, \methodname takes masked speech token sequence $\bm{x}_t$ as input and decomposes it into low-level component $\bm{x}_t^{\text{low}} = \bm{x}^{r_1:r_2}_t$ and high-level component $\bm{x}_t^{\text{high}} = \bm{x}^{r_3:r_{12}}_t$. Then, according to the inherent hierarchy of speech tokens, we disentangle the visual features extraction from the input video $\mathcal{V}$ and inject them into \methodname. The lip motion features $\bm{c}_\text{lip}$ and identity features $\bm{c}_\text{id}$ are embedded into the low-level blocks to refine the generation of content- and timbre-centric tokens, while the emotional expression features $\bm{c}_\text{emo}$ are injected into the high-level blocks for enhancing prosody-related tokens generation. 
Finally, \methodname outputs concrete scores for the reverse diffusion process to recover the masked tokens, which are decoded by the codec to synthesize high-fidelity speech.

\subsection{Disentangled Visual Conditioning}
\label{subsec:f2a}
\paragraph{Lip adapter for content modelling.}
Due to the strong temporal alignment between lip motion and speech content~\cite{DBLP:conf/cvpr/DaiCDWCW024}, we extract visual features using AV-HuBERT~\cite{avhubert:conf/iclr/ShiHLM22}, taking the last-layer hidden states as they encode the most discriminative audio-visual semantics. These features are projected via a Multilayer Perceptron (MLP) to obtain $\bm{c}_\text{lip} \in \mathbb{R}^{L \times C}$, where $L$ and $C$ denote the sequence length and channel dimension, respectively, matching those of the masked low-level speech tokens $\bm{m}_t^{\text{low}}$. 

\paragraph{Identity adapter for timbre modelling.} 
Since both speech timbre and facial appearance encode speaker identity—despite lacking direct correspondence, we align their representation through cross-modal identity modelling. Specifically, visual identity features are extracted from facial images using ArcFace~\cite{arcface:journals/pami/DengGYXKZ22} and projected via an MLP into $\bm{c}_\text{id} \in \mathbb{R}^{L \times C_{\text{ge2e}}}$, where $C_{\text{ge2e}}$ matches the channel dimension of acoustic identity features extracted by the GE2E model~\cite{GE2E/WanWPL18}. The two modalities are aligned by minimizing the $\ell_1$ distance between their embeddings. The $\bm{c}_\text{id}$ is then fed into an MLP to generate modulation parameters for our dual-level AdaLN for timbre conditioning.

\paragraph{Emotion adapter for prosody modelling.}
Prosody refers to the non-lexical acoustic properties that convey the emotion of the speaker~\cite{frick1985communicating}. We leverage facial expression as a proxy signal by employing Poster2~\cite{posterv2:journals/corr/abs-2301-12149}, which is a strong video facial expression recognition model. To suppress identity-biased fluctuations, we only predict emotional class over all frames and apply temporal smoothing over 0.5-second windows, reducing the sequence to length $L^{\text{emo}}$. A learnable embedding layer then maps the smoothed class sequence to emotional features $\bm{c}_\text{emo} \in \mathbb{R}^{L_{\text{emo}} \times C}$, conditioning high-level speech tokens to modulate prosody.

\begin{figure*}[t]
    \centering
    \includegraphics[width=0.99\linewidth]{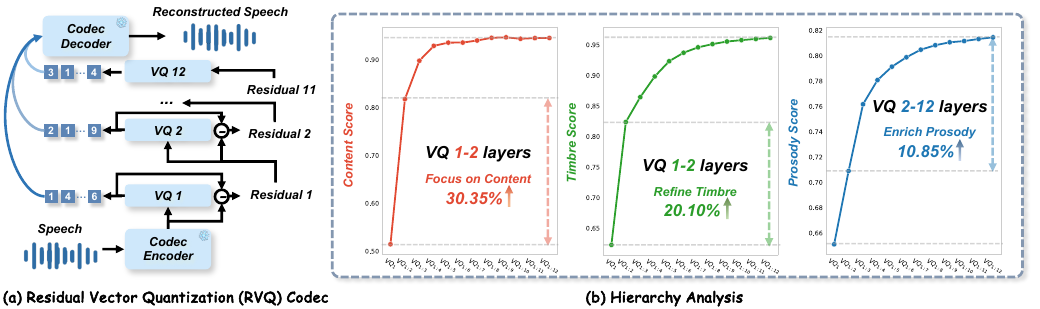}
    \caption{
    \textbf{Hierarchy analysis of speech token.} (a) RVQ codec encodes and decodes speech through multiple VQ layers. (b) The x-axis denotes cumulative decoding across token levels, and the y-axis reports scores for semantic fidelity, timbre similarity, and prosody quality. It can be observed that speaker-aware semantics improvements concentrate in lower layers, while prosody gains emerge in higher layers.
    }
    \label{fig:rvq}
\end{figure*}

\subsection{Hierarchical Masked Token Prediction}
\label{subsec:transformer}
\paragraph{Hierarchical speech tokenization and diffusion.}
Given a single-channel speech signal, we utilize the RVQ-based codec~\cite{maskgct:journals/corr/abs-2409-00750} to compresses it into tokens represented as $\bm{x}^{r_1:r_{12}} = \{1,\ldots,C_\text{code}\}^{12\times L}$, where $r_i$ is the $i$-th level of token, $L$ is the length of the token sequence, respectively. The number of RVQ layers is 12 with a codebook size $C_\text{code}=1,024$ in each level. We partition RVQ tokens into low-level $\bm{x}_t^{\text{low}} = \bm{x}^{r_1:r_2}_t$ and high-level $\bm{x}_t^{\text{high}} = \bm{x}^{r_3:r_{12}}_t$, reflecting the hierarchical structure of speech and consistent with the hierarchy analysis in Figure~\ref{fig:rvq}. The tokens are then masked via the discrete diffusion process of SEDD~\cite{SEDD:conf/icml/LouME24}, as formalized in~\cref{eq:delta_transition}, yielding $\bm{x}_t^{\text{low}}$ and $\bm{x}_t^{\text{high}}$ at step $t$.

\paragraph{Hierarchical codec diffusion transformer.}
The proposed \methodname serves as the score network in~\cref{eq:score_entropy}, predicting concrete scores for masked speech tokens that parametrize the transition rate from the masked state to each valid token. 
To align visual cues with the hierarchical structure of speech, we employ two complementary conditioning mechanisms: (i) direct concatenation for fine-grained and frame-synchronized signals such as lip motion, and (ii) dual-scale AdaLN for class-like attributes like speaker identity and emotion. 
For the content conditioning, the masked features $\bm{m}_t^{\text{low}} \in \mathbb{R}^{L \times C}$ are first concatenated with lip motion features $\bm{c}_\text{lip}$ along the channel dimension, followed by a linear layer to enhance temporally synchronized fusion. 
For the timbre conditioning, we utilize a MLP predicts channel-level scale and shift parameters $\alpha_\text{id}^1,\gamma_\text{id}^1,\beta_\text{id}^1,\alpha_\text{id}^2,\gamma_\text{id}^2,\beta_\text{id}^2\in\mathbb{R}^{C}$ based on both identity features $\bm{c}_\text{id}$ and time $t$ features. We can formulate the identity conditioning of the single-scale AdaLN as:
\begin{equation}
\label{eq:channellevel}
(1+\gamma_\text{id}^i)\cdot\frac{\bm{h}_t -\mu(\bm{h}_t )}{\sigma(\bm{h}_t )}+\beta_\text{id}^i,
\end{equation}
where $i=\{1,2\}$ for multi-head attention and feed-forward network, and $\bm{h}_t \in \mathbb{R}^{L\times C}$ is the hidden embedding after layer normalization in low-level blocs. $\mu(\cdot)$ and $\sigma(\cdot)$ are the mean and standard deviation for $\bm{h}_t$ across the channel dimension. 
Furthermore, for the prosody conditioning, we introduce a temporal MLP to predict \textbf{t}emporal-level scale parameters $\gamma_\text{emo,t}^1, \gamma_\text{emo,t}^2\in\mathbb{R}^{L^{\text{emo}}}$ using emotion features $\bm{c}_{\text{emo}}$ and time $t$ features, and a channel MLP to predict \textbf{c}hannel-level scale and shift parameters $\alpha_\text{emo,c}^1,\gamma_\text{emo,c}^1,\beta_\text{emo,c}^1,\alpha_\text{emo,c}^2,\gamma_\text{emo,c}^2,\beta_\text{emo,c}^2\in\mathbb{R}^{C}$ using pooling emotion features and time features. We can formulate the prosody conditioning of the dual-scale AdaLN as:
\begin{equation}
\label{eq:duallevel}
\underbrace{\gamma_\text{emo,t}^i\otimes \mathbf{1}_{25}}_\text{Temporal-level} \cdot \underbrace{\left((1+\gamma_\text{emo,c}^i)\cdot\frac{\bm{h}_t-\mu(\bm{h}_t)}{\sigma(\bm{h}_t)}+\beta_\text{emo,c}^i\right)}_\text{Channel-level},
\end{equation}
where $i=\{1,2\}$, $\otimes$ denotes Kronecker product, and $\mathbf{1}_{25}\in\mathbb{R}^{25}$ is an all-ones vector to up-sample $\gamma_\text{te}^i$ with $L_{\text{emo}}=\frac{L}{25}$ parameters to align with the hidden embedding with 50 Hz sampling rate. Finally, for the output, we incorporate 12 linear score heads to predict concrete scores for each level. 
These conditioning mechanisms enable \methodname to faithfully modulate speech generation according to the cross-modal prior, bridging the gap between video and speech.

\subsection{Training and Inference}
\label{subsec:training}
\paragraph{Training.} 
\methodname is optimized by multi-level DSE loss based on~\cref{eq:score_entropy} with the sum across all 12 RVQ levels as $\mathcal{L}_\text{score} = \sum_{i=1}^{12}\mathcal{L}_\text{DSE}(\bm{x}^{r_i},t,\bm{c})$. For conducting predictor-free guidance, we randomly set $\varnothing$ with 10\% probability for each condition and enforce all conditions set to $\varnothing$ for 10\% samples. An additional loss $\mathcal{L}_\text{id}=\ell_1(\bm{c}_\text{id},\bm{c}_\text{GE2E})$ aligns the visual identity embedding $\bm{c}_\text{id}$ with the GE2E speech embedding $\bm{c}_\text{GE2E}$ to reinforce speaker consistency. To summarize, the total loss function $\mathcal{L}_\text{total}$ is defined as follows:
\begin{equation}
\label{eq:totalloss}
\mathcal{L}_\text{total} = \mathcal{L}_\text{score} + \lambda \mathcal{L}_\text{id}, 
\end{equation}
where $\lambda$ is set to 100.0 in our experiments. 

\paragraph{Inference.}
Following~\cref{eq:backward}, the reverse process is executed with Euler sampling~\cite{SEDD:conf/icml/LouME24} and enhanced predictor-free guidance~\cite{demoface/abs-2502-01046} with 64 sampling steps. 
Notably, to ensure training stability, we utilize ground truth acoustic features to replace $\bm{c}_\text{id}$ and $\bm{c}_\text{emo}$ during training, whereas only visual features are used during inference.

\section{Experimental Results}
\label{sec:result}

\subsection{Experimental Setups}
\paragraph{Datasets.} 
Our \methodname is trained on the VoxCeleb2~\cite{chung2018voxceleb2} dataset, which provides large-scale speaker-diverse audiovisual recordings. 
To ensure well-aligned data, we perform a multi-stage data preprocessing pipeline. We first resample all audio to 16 kHz and employ a speech language identification model~\cite{speechbrain, valk2021slt} to filter out non-English utterances. We then apply a speaker diarization model~\cite{Plaquet23} to remove multi-speaker segments, followed by the ClearerVoice speech separation model~\cite{ClearerVoice:journals/corr/abs-2506-19398} to enhance signal-noise ratio. Finally, we leverage~\cite{whisper/RadfordKXBMS23} to discard misaligned text–speech pairs. 
Finally, the pre-processing dataset comprises 261.5 hours of audio recordings with 169k utterances across 7 basic emotions and 3,438 speakers. 
For evaluation, we test our models on two in-the-wild datasets without any specific training, LRS2~\cite{LRS2/cvpr/SonChung} and LRS3~\cite{LRS3/abs-1809-00496}.

\begin{table*}[thbp]
\setlength\tabcolsep{4.5pt}
\footnotesize
\centering
\begin{tabular}{rccccccccccc}
\toprule
\multirow{2}{*}{Methods} & \multirow{2}{*}{Venue} & \multirow{2}{*}{A} & \multirow{2}{*}{V} & \multicolumn{4}{c}{\textit{Naturalness}} & \multicolumn{2}{c}{\textit{Synchronization}} & \multicolumn{2}{c}{\textit{Expressiveness}} \\
 & & & & WER$\downarrow$ & DNSMOS$\uparrow$ & UTMOS$\uparrow$ & MCD$\downarrow$ & LSE-C$\uparrow$ & LSE-D$\downarrow$ & EmoAcc$\uparrow$ & SpkSim$\uparrow$ \\ \hline
\specialrule{0em}{2pt}{1pt}
Ground Truth & - & - & - & 2.29 & 3.29 & 3.57 & 0.00 & 6.66 & 6.89 & 100.00 & 1.0000 \\ 
\hdashline
\specialrule{0em}{1pt}{1pt}
Lip2Wav$^\dag$~\cite{lip2wav:conf/cvpr/PrajwalMNJ20} & CVPR'20 & \ding{51} & \ding{51} & 98.68 & 2.47 & 1.29 & 13.43 & 3.37 & 9.85 & 63.11 & 0.4785 \\ 
MTL~\cite{MTL_lip:conf/icassp/KimHR23} & ICASSP'23 & \ding{51} & \ding{51} & 76.61 & 2.42 & 1.28 & 9.84 & 5.87 & 7.51 & 61.24 & 0.3347 \\ 
EmoDubber$^\dag$~\cite{EmoDubber:conf/cvpr/0001P0QPH0H25} & CVPR'25 & \ding{51} & \ding{51}  & 41.52 & 2.95 & 2.83 & \cellcolor{lighterblue}{9.25} & 6.88 & 6.85 & 72.01 & \cellcolor{lighterblue}{0.6052} \\ 
\specialrule{0em}{1pt}{1pt}
\hline
\specialrule{0em}{1pt}{1pt}
DiffV2S~\cite{diffv2s:conf/iccv/ChoiHR23} & ICCV'23 & \ding{55} & \ding{51} & 41.07 & 2.56 & 3.06 & - & - & - & - & - \\ 
LTBS$^\dag$~\cite{LTBS:conf/aaai/KimKC24} & AAAI'24 & \ding{55} & \ding{51} & 84.00 & 2.36 & 2.42 & - & - & - & - & - \\ 
AlignDiT~\cite{aligndit:journals/corr/abs-2504-20629} & ACM MM'25 & \ding{55} & \ding{51} & 31.37 & 3.24 & 3.76 & 10.02 & 6.95 & 6.82 & 76.11 & 0.5597 \\ 
FTV~\cite{FTV:conf/cvpr/KimCKJC25} & CVPR'25 & \ding{55} & \ding{51} & 30.37 & 3.22 & \cellcolor{lightblue}\textbf{3.99} & 10.54 & {7.08} & {6.66} & 73.19 & 0.5981 \\ 
\specialrule{0em}{1pt}{1pt}
\hline
\specialrule{0em}{1pt}{1pt}
\multirow{2}{*}{\textbf{\methodname{}$^\dag$}(\textbf{ours})} & \multirow{2}{*}{-} & \ding{55} & \ding{51} & \cellcolor{lighterblue}{29.41} & \cellcolor{lightblue}\textbf{3.50} & \cellcolor{lighterblue}\textbf{3.84} & 9.62 & \cellcolor{lightblue}\textbf{7.15} & \cellcolor{lightblue}\textbf{6.58} & \cellcolor{lightblue}\textbf{79.41} & 0.5678 \\ 
 &  & \ding{51} & \ding{51} & \cellcolor{lightblue}\textbf{28.98} & \cellcolor{lighterblue}{3.44} & {3.80} & \cellcolor{lightblue}\textbf{8.69} & \cellcolor{lighterblue}{7.10} & \cellcolor{lighterblue}{6.61} & \cellcolor{lighterblue}{77.08} & \cellcolor{lightblue}\textbf{0.6715} \\ 
\bottomrule
\end{tabular}
\caption{\textbf{Quantitative results on LRS3.} A/V indicate use of audio/video guidance (\ding{51}/\ding{55}). The superscript $^\dag$ indicates that the model is not trained on LRS3. $\uparrow$ ($\downarrow$) indicates that higher (lower) is better. Best results are highlighted in \highlightblue{\textcolor{black}{\textbf{deeper blue}}}, second-best in \highlightlightblue{\textcolor{black}{lighter blue}}.
\label{tab:lrs3}}
\vspace{-2mm}
\end{table*}

\begin{table*}[thbp]
\setlength\tabcolsep{4.5pt}
\footnotesize
\centering
\begin{tabular}{rccccccccccc}
\toprule
\multirow{2}{*}{Methods} & \multirow{2}{*}{Venue} & \multirow{2}{*}{A} & \multirow{2}{*}{V} & \multicolumn{4}{c}{\textit{Naturalness}} & \multicolumn{2}{c}{\textit{Synchronization}} & \multicolumn{2}{c}{\textit{Expressiveness}} \\
 & & & & WER$\downarrow$ & DNSMOS$\uparrow$ & UTMOS$\uparrow$ &  MCD$\downarrow$ & LSE-C$\uparrow$ & LSE-D$\downarrow$ & EmoAcc$\uparrow$ & SpkSim$\uparrow$ \\ \hline
\specialrule{0em}{2pt}{1pt}
Ground Truth & - & - & - & 8.93 & 3.14 & 3.05 & 0.00 & 7.20 & 6.67 & 100.00 & 1.0000\\ 
\hdashline
\specialrule{0em}{1pt}{1pt}
Lip2Wav$^\dag$~\cite{lip2wav:conf/cvpr/PrajwalMNJ20} & CVPR'20 & \ding{51} & \ding{51} & 100.05 & 2.47 & 1.31 & 14.09 & 3.83 & 9.80 & 54.38 & 0.4438 \\ 
MTL~\cite{MTL_lip:conf/icassp/KimHR23} & ICASSP'23 & \ding{51} & \ding{51} & 58.03 & 2.42 & 1.30 & 10.71 & 6.58 & 7.16 & 63.89 & 0.3556 \\ 
EmoDubber~\cite{EmoDubber:conf/cvpr/0001P0QPH0H25} & CVPR'25 & \ding{51} & \ding{51}  & 47.60 & 2.84 & 2.77 & \cellcolor{lightblue}\textbf{7.02} & 7.42 & 6.60 & 66.76 & 0.5252 \\ 
\specialrule{0em}{1pt}{1pt}
\hline
\specialrule{0em}{1pt}{1pt}
DiffV2S~\cite{diffv2s:conf/iccv/ChoiHR23} & ICCV'23 & \ding{55} & \ding{51 }& 54.86 & 2.36 & 2.95   & - & - & - & - & - \\ 
LTBS$^\dag$~\cite{LTBS:conf/aaai/KimKC24} & AAAI'24 & \ding{55} & \ding{51} & 94.30 & 2.17 & 2.29 & - & - & - & - & - \\ 
AlignDiT$^\dag$~\cite{aligndit:journals/corr/abs-2504-20629} & ACM MM'25 & \ding{55} & \ding{51} & 42.26 &  3.13  & 3.65 & 8.46 & 7.50 & 6.58 & 67.01 & 0.5187 \\ 
FTV~\cite{FTV:conf/cvpr/KimCKJC25} & CVPR'25 & \ding{55} & \ding{51} & \cellcolor{lightblue}\textbf{38.09} & 3.11 & \cellcolor{lightblue}\textbf{3.88} & 12.91 & {7.71} & {6.35} & \cellcolor{lighterblue}{67.84} & \cellcolor{lighterblue}{0.5368}\\ 
\specialrule{0em}{1pt}{1pt}
\hline
\specialrule{0em}{1pt}{1pt}
\multirow{2}{*}{\textbf{\methodname{}$^\dag$}(\textbf{ours})} & \multirow{2}{*}{-} & \ding{55} & \ding{51} & \cellcolor{lighterblue}{39.99} & \cellcolor{lightblue}\textbf{3.35} & \cellcolor{lighterblue}{3.68} & 8.74 & \cellcolor{lightblue}\textbf{7.95} & \cellcolor{lightblue}\textbf{6.17} & \cellcolor{lightblue}\textbf{68.21} & 0.5222 \\ 
 &  & \ding{51} & \ding{51} & {40.75} & \cellcolor{lighterblue}{3.27} & 3.38 & \cellcolor{lighterblue}{8.36} & \cellcolor{lighterblue}{7.83} & \cellcolor{lighterblue}{6.24} & 65.65 & \cellcolor{lightblue}\textbf{0.5954} \\ 
\bottomrule
\end{tabular}
\caption{\textbf{Quantitative results on LRS2.}  A/V indicate use of audio/video guidance (\ding{51}/\ding{55}). The superscript $^\dag$ indicates that the model is not trained on LRS2. $\uparrow$ ($\downarrow$) indicates that higher (lower) is better. Best results are highlighted in \highlightblue{\textcolor{black}{\textbf{deeper blue}}}, second-best in \highlightlightblue{\textcolor{black}{lighter blue}}.\label{tab:lrs2}}
\vspace{-2mm}
\end{table*}

\paragraph{Evaluation metrics.}
The generation performance is evaluated using both subjective and objective metrics. 
For subjective assessment, we conduct a Mean Opinion Score (MOS) and A/B testing. 
For objective assessment, we first quantify spectral differences with Mel Cepstral Distortion (MCD)~\cite{visualvoicecloning/ChenTQZLW22}, DNSMOS~\cite{dnsmos:conf/icassp/ReddyGC21}, and UTMOS~\cite{utmos}, which are widely used networks to estimate perceptual audio quality. We also calculate the Word Error Rate (WER)~\cite{WER/WangSZRBSXJRS18,whisper/RadfordKXBMS23} to gauge intelligibility. 
For the synchronization, we report the distance and confidence scores of lip sync errors (LSE-C and LSE-D) between speech and video using the pre-trained SyncNet~\cite{syncnet:conf/accv/ChungZ16a}. 
For the expressiveness, we calculate cosine similarity metrics based on ECAPA-TDNN~\cite{DBLP:conf/interspeech/DesplanquesTD20} to obtain speaker identity similarity (SpkSim). Additionally, we evaluate emotion accuracy (EmoAcc) using a strong speech emotion recognition model~\cite{emo2vec:conf/acl/MaZYLGZ024,emobox:conf/interspeech/MaCZZCLY0H24}.

\begin{figure*}[t]
    \centering
    \includegraphics[width=0.97\linewidth]{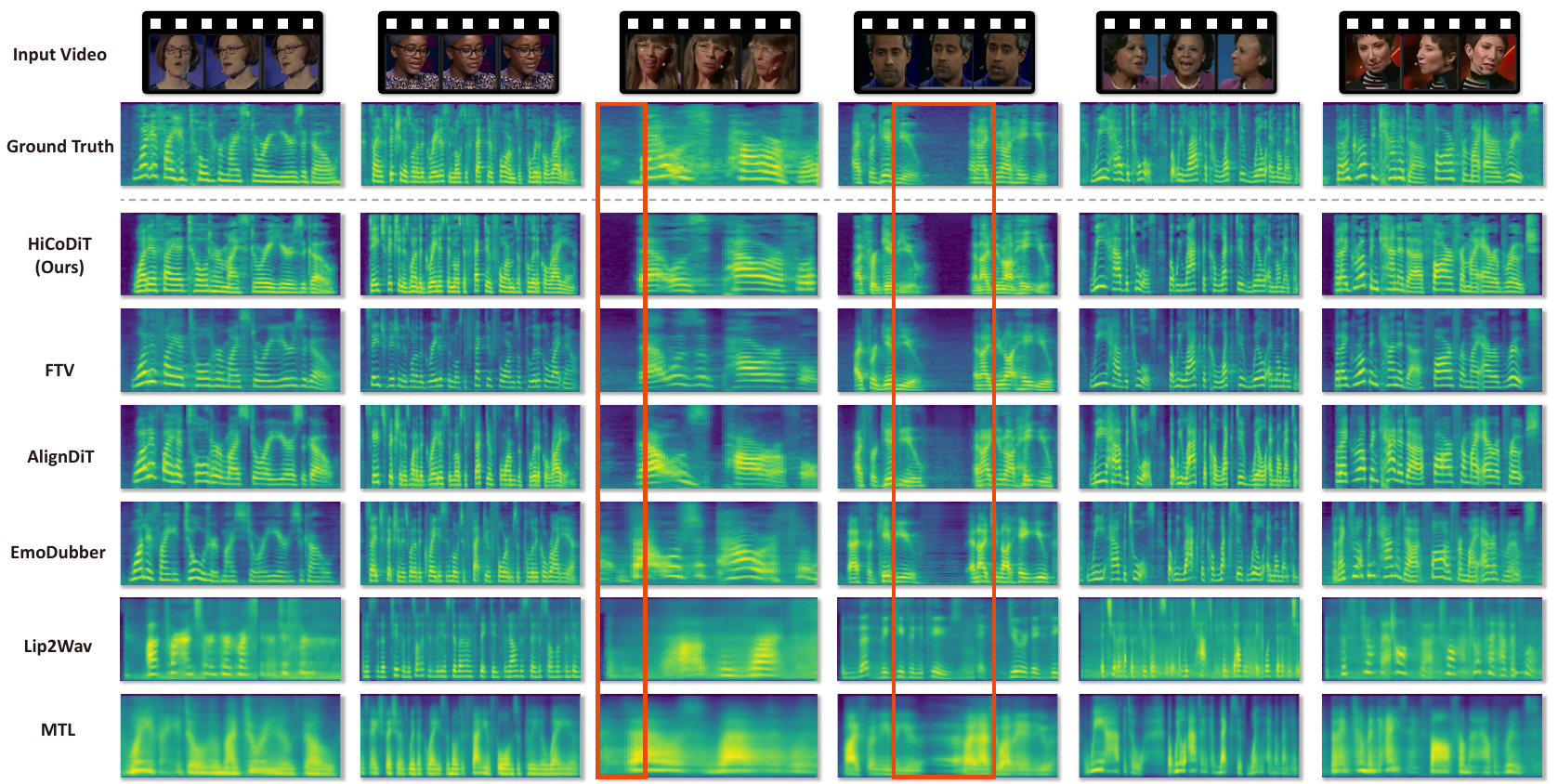}
    \caption{\textbf{The visualization of the mel-spectrograms of ground truth (GT) and synthesized speech obtained by different models.} As highlighted in the red boxes, the spectrograms generated by our method exhibit higher clarity with improved signal-to-noise ratio.}
    \label{fig:mel}
\end{figure*}

\paragraph{Implementation details.}
For the speech tokenization, we employ a pre-trained RVQ-based codec from MaskGCT~\cite{maskgct:journals/corr/abs-2409-00750}, and adopt a log-linear noise schedule $\sigma(t)$~\cite{SEDD:conf/icml/LouME24} for the diffusion process, where the expectation of the number of masked tokens is linear with time $t$. 
For the disentangled visual conditioning, AV-HuBERT-Large, ArcFace, and Poster2 are used for lip, identity, and emotion feature extraction, respectively. 
For the transformer, the numbers of low- and high-level blocks are 8 and 8, respectively. The channel dimension $C$ is set to 768 with 12 attention heads. 
During training, we use the AdamW optimizer~\cite{adamw/LoshchilovH19} with a learning rate of 1e-4, batch size 32. The total number of iterations is 200k. 
During inference, we employ an Euler sampler to perform the reverse process in 64 steps. For multi-conditional guidance, we adopt the enhanced predictor-free guidance~\cite{demoface/abs-2502-01046}, setting the joint guidance scale to $w_{\text{all}}=2.5/2.25$ and the compositional scales to $w_{\text{id}}=1.25/1.25$, $w_{\text{id}}=1.5/1.5$, and $w_{\text{lip}}=2.0$ for the LRS3 and LRS2 datasets, respectively.

\paragraph{Baseline models.}
Our method is compared with several state-of-the-art approaches: FTV~\cite{FTV:conf/cvpr/KimCKJC25}, AlignDiT~\cite{aligndit:journals/corr/abs-2504-20629}, EmoDubber~\cite{EmoDubber:conf/cvpr/0001P0QPH0H25}, MTL~\cite{MTL_lip:conf/icassp/KimHR23}, Lip2Wav~\cite{lip2wav:conf/cvpr/PrajwalMNJ20}, LTBS~\cite{LTBS:conf/aaai/KimKC24}, and DiffV2S~\cite{diffv2s:conf/iccv/ChoiHR23}. 
For FTV, the test samples on both LRS2 and LRS3 are provided by the authors.
For AlignDiT, MTL, and Lip2Wav, we use the publicly released models for inference.
For EmoDubber, we reproduce results using the official training code.
Since no public models or test samples are available, we report results as cited in their original publications for both LTBS and DiffV2S.

\subsection{Quantitative Evaluation}

\paragraph{Objective evaluation.} 
Tables~\ref {tab:lrs3} and~\ref{tab:lrs2} summarize the objective evaluation results on the LRS3 and LRS2 datasets, respectively. Although our \methodname is not trained on either LRS3 or LRS2, it achieves leading performance on key metrics, including overall speech quality (UTMOS, DNSMOS), intelligibility (WER), and lip synchronization (LSE-C). While EmoDubber achieves the best spectral clarity on MCD by directly optimizing spectrograms, our method, which focuses on discrete speech token generation, achieves the second-best performance. Furthermore, our method exhibits a degradation in speaker similarity relative to FTV, reflecting the limited diversity of our training data. However, when a speech signal is introduced as identity guidance, our method achieves the highest score on this metric, showing great voice cloning ability. Similar trends are observed on LRS2. Overall, these results demonstrate the effectiveness of our hierarchical masked token prediction for VTS.

\paragraph{Subjective evaluation.}
We further conduct the subjective evaluation with 20 participants, to compare our \methodname with SOTA methods.
Specifically, we introduce five MOS with rating scores from 1 to 5 in 0.5 increments, including $\text{MOS}_\text{nat}$, $\text{MOS}_\text{exp}, \text{MOS}_\text{syn}$ for speech naturalness, expressiveness, and lip-synchronization. We randomly generate 30 samples from the test set. The scoring results of the user study are presented in Table~\ref{tab:userstudy}, demonstrating that \methodname outperforms SOTA methods across nearly all metrics, particularly surpassing  2.94\%in $\text{MOS}_\text{syn}$ with the ground truth. 
Tables~\ref {tab:lrs3} and~\ref{tab:lrs2} summarize the objective evaluation results on the LRS3 and LRS2 datasets, respectively. Although our \methodname is not trained on either LRS3 or LRS2, it achieves leading performance on key metrics, including overall speech quality (UTMOS, DNSMOS), intelligibility (WER), and lip synchronization (LSE-C). While EmoDubber achieves the best spectral clarity on MCD by directly optimizing spectrograms, our method, which focuses on discrete speech token generation, achieves the second-best performance. Furthermore, our method exhibits a degradation in speaker similarity relative to FTV, reflecting the limited diversity of our training data. However, when a speech signal is introduced as identity guidance, our method achieves the highest score on this metric, showing great voice cloning ability. Similar trends are observed on LRS2. Overall, these results demonstrate the effectiveness of our hierarchical masked token prediction for VTS.
Furthermore, the proposed model HiCoDiT achieves the highest $\text{MOS}_\text{nat}$ (3.17) and $\text{MOS}_\text{sync}$  (3.50), indicating superior naturalness and synchronization compared to existing methods like AlignDiT and FTV. Although the expressiveness is slightly lower than FTV, indicating that a more diverse speaker dataset can enhance expressiveness.

\begin{table}[h]
\setlength\tabcolsep{5.0pt}
\small
\centering
\begin{tabular}{rccc}
\toprule
Methods & $\text{MOS}_\text{nat}\uparrow$ & $\text{MOS}_\text{exp}\uparrow$ & $\text{MOS}_\text{syn}\uparrow$ \\ \hline
\specialrule{0em}{1.5pt}{1.5pt}
Ground Truth   & 3.07$\pm$1.02 & 3.30$\pm$1.19 & 3.40$\pm$0.93   \\
\hdashline
\specialrule{0em}{1.5pt}{1.5pt}
AlignDiT~\cite{aligndit:journals/corr/abs-2504-20629}   & 2.47$\pm$1.19 & 2.63$\pm$1.30 & 3.13$\pm$0.75   \\
FTV~\cite{FTV:conf/cvpr/KimCKJC25}   & 2.80$\pm$1.03 & \textbf{2.90$\pm$1.45} & 3.48$\pm$1.02  \\
\specialrule{0em}{1.pt}{1.0pt}
\hline
\specialrule{0em}{1.pt}{1.0pt}
\methodname (\textbf{ours})   & \textbf{3.17$\pm$1.31} & 2.88$\pm$1.53 & \textbf{3.50$\pm$0.86} \\
\bottomrule
\end{tabular}
\caption{\textbf{Subjective evaluation} on speech naturalness, expressiveness, and synchronization, compared with other SOTA methods. \label{tab:userstudy}}
\vspace{-0.5mm}
\end{table}

\begin{table}[htbp]
\centering
\setlength\tabcolsep{5.0pt}
\footnotesize
{\begin{tabular}{rccc}
\toprule
\textbf{A vs. B} & \textbf{A wins (\%)} & \textbf{Neutral} & \textbf{B wins (\%)} \\
\midrule
Ours vs. AlignDiT~\cite{aligndit:journals/corr/abs-2504-20629} & \textbf{57.0} & 4.9 & 38.1 \\
Ours vs. FTV~\cite{FTV:conf/cvpr/KimCKJC25} & \textbf{52.1} & 6.1 & 41.8 \\
GT vs. FTV~\cite{FTV:conf/cvpr/KimCKJC25} & \textbf{51.5} & 14.0 & 34.5 \\
GT vs. Ours &  45.5 & 0.6 & \textbf{53.9} \\
\bottomrule
\end{tabular}}
\caption{\textbf{A/B testing results.}  We report the preferences (\%) between A and B across various aspects of synthesized speech. \label{tab:ab_test_naturalness}}
\vspace{-2mm}
\end{table}

In addition, the Table~\ref{tab:ab_test_naturalness} compares A/B test preferences for synthesized speech. Our method demonstrates clear superiority over AlignDiT, achieving a 57.0\% preference, and also outperforms FTV with 52.1\% preference. 
Additionally, ground-truth speech is preferred over FTV (51.5\%) and is outperformed by our method with a 53.9\% preference, showing the strength of our model in generating high-quality speech nearly indistinguishable from real speech.

\begin{table*}[thbp]
\setlength\tabcolsep{4.5pt}
\footnotesize
\centering
\begin{tabular}{clcccccccc}
\toprule
Datasets & Ablations & WER$\downarrow$ & DNSMOS$\uparrow$ & UTMOS$\uparrow$ & MCD$\downarrow$ & LSE-C$\uparrow$ & LSE-D$\downarrow$ & EmoAcc$\uparrow$ & SpkSim$\uparrow$ \\
\specialrule{0em}{1.pt}{1.0pt}
\hline
\specialrule{0em}{1.pt}{1.0pt}
\multirow{3}{*}{LRS3}
& w/o Hierarchical Modeling & 30.65 & 3.36 & 3.73 & 10.07 & 7.02 & 6.75 & 76.98 & 0.5652 \\
& w/o Dual Scale AdaLN  & 29.60 & 3.45 & \textbf{3.92} & 9.75 & 7.12 & 6.60 & 78.55 & 0.5621 \\
\specialrule{0em}{1.pt}{1.0pt}
& \textbf{\methodname (full)} & \textbf{29.41} & \textbf{3.50} & {3.84} & \textbf{9.62} & \textbf{7.15} & \textbf{6.58} & \textbf{79.41} & \textbf{0.5678} \\
\specialrule{0em}{1.pt}{1.0pt}
\hline
\specialrule{0em}{1.pt}{1.0pt}
\multirow{3}{*}{LRS2}
& w/o Hierarchical Modeling & 44.57 & 3.18 & 3.48 & 9.43 & 7.66 & 6.47 & 64.69 & 0.4946 \\
& w/o Dual Scale AdaLN & 41.01 & 3.30  & \textbf{3.75} & 9.33 & 7.88 & 6.22 & \textbf{68.61} & 0.5155 \\
\specialrule{0em}{1.pt}{1.0pt}
& \textbf{\methodname (full)} &  \textbf{39.99} & \textbf{3.35} & {3.68} & \textbf{8.74} & \textbf{7.95} & \textbf{6.17} & {68.21} & \textbf{0.5222} \\ 
\bottomrule
\end{tabular}
\caption{\textbf{Ablation study on LRS3 and LRS2.} Best results are highlighted in {\textcolor{black}{\textbf{Bold}}}.}
\label{tab:ablation_lrs3_lrs2}
\end{table*}

\subsection{Qualitative Results}
\paragraph{Qualitative spectrogram comparisons.}
As shown in Figure~\ref{fig:mel}, we compare generated mel-spectrograms with other methods. For Lip2Wav and MTL, we observe severe over-smoothing or acoustic artifacts, resulting in significant degradation of speech quality and limiting their practical utility, which may be attributed to the insufficient probabilistic modeling capacity of the generative models used. 
Methods based on powerful diffusion models, produce high-quality speech. However, their mel-spectrograms still exhibit noise in the silent clip. In contrast, our method generates clarity mel-spectrograms with richer acoustic details and precise lip-synchronization, benefiting from the strong speech reconstruction capability of Codec.

\subsection{Ablation Studies}
\paragraph{Ablation on hierarchical modeling.}
We explore the impact of hierarchical modeling on video-to-speech generation in Table~\ref{tab:ablation_lrs3_lrs2}. The removal of hierarchical modeling collapses the multi-level speech representation into a single uniform module, while simultaneously forcing visual conditioning to be injected across all tokens. Experimental results demonstrate that performance degrades significantly across all metrics, underscoring the validity of our proposed speech hierarchy prior. This further indicates that visual features corresponding to specific attributes should align with speech tokens carrying matching content. 

\paragraph{Ablation on dual scale AdaLN.}
To demonstrate the effectiveness of the proposed Dual-scale AdaLN, we utilize the vanilla adaLN of DiT~\cite{DiT:conf/iccv/PeeblesX23} and replace the temporal embedding with utterance-level emotional embedding combined with global style as an acoustic guidance. 
As shown in Table~\ref{tab:ablation_lrs3_lrs2}, pooling dynamic emotions struggles to model prosody dynamics, with a negligible decrease in terms of EmoAcc, while other metrics gain a lot. The results highlight the effectiveness of our dual-scale AdaLN mechanism.  
\begin{figure}[t]
    \centering
    \vspace{-0.15cm}
    \includegraphics[width=1.0\linewidth]{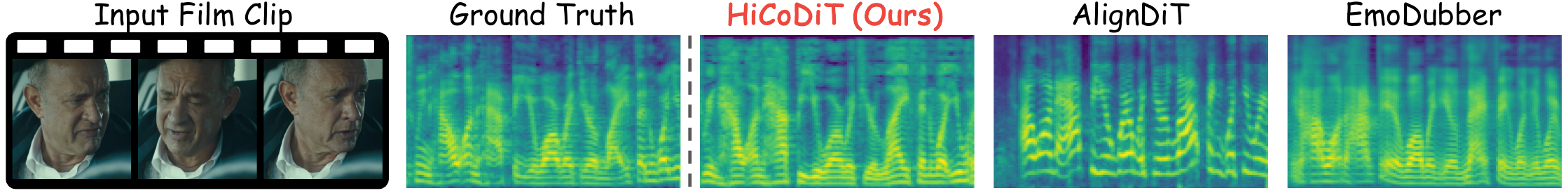}
        \vspace{-0.22cm}
    \scriptsize
    \caption{Comparison of generated Mels on real-world film data.}
    \label{fig:ood}
    \vspace{-0.2cm}
\end{figure}

\begin{table}[t]
\small

\resizebox{\columnwidth}{!}{
    \begin{tabular}{lccccccc}
    \toprule
    \textbf{Method} & \textbf{WER}$\downarrow$ & \textbf{MCD}$\downarrow$ & \textbf{DNSMOS}$\uparrow$ &  \textbf{Emo}$\uparrow$ & \textbf{Spk}$\uparrow$ & \textbf{LSE-D}$\downarrow$  \\ \hline
    \specialrule{0em}{1.pt}{0.5pt}
    EmoDubber   & 88.3 & 9.9  & 2.8 & 76.5  & 45.1 & 7.72 \\ 
    AlignDiT   & 80.8 & 11.4  & 3.2 & 75.2  & \textbf{58.5} & 8.23 \\ 
    \specialrule{0em}{1.pt}{0.5pt}
    \hline
    \specialrule{0em}{1.pt}{0.5pt}
    \textbf{\methodname} & \textbf{58.7} & \textbf{9.8}  & \textbf{3.5} & \textbf{82.0}  & 50.1 & \textbf{7.60} \\ 
     \bottomrule
    \end{tabular}
    }
\vspace{-0.1cm}
\caption{Quantitative comparison of real-world OOD film data.}
\label{tab:ood}
\vspace{-0.11cm}
\end{table}

\paragraph{Ablation on out-of-domain data.}
To assess generalization in complex, real-world environments, we curate an authentic film benchmark comprising 160 utterances across 56 speakers from \textit{CinePile} to ensure realistic audio-visual complexity. We compared \methodname against primary open-source SOTA methods EmoDubber and AlignDiT. Table~\ref{tab:ood} and Figure~\ref{fig:ood} demonstrate that our method achieves robust intelligibility and lip-synchronization on this challenging OOD data, underscoring \methodname's robustness and adaptability to authentic scenarios.

\begin{table}[t]
    \small
    \resizebox{\columnwidth}{!}{
        \begin{tabular}{lccccccc}
        \toprule
        \textbf{Ablations} & \textbf{WER}$\downarrow$ & \textbf{MCD}$\downarrow$ & \textbf{DNSMOS}$\uparrow$ &  \textbf{Emo}$\uparrow$ & \textbf{Spk}$\uparrow$ & \textbf{LSE-D}$\downarrow$  \\ \hline
        \specialrule{0em}{1.pt}{0.5pt}
        (a) wo GE2E $\mathcal{L}_{\text{id}}$   & \textbf{29.38} & 10.18  & 3.41 & 74.47  & 34.10 & 6.71 \\ 
        (b) wo Poster2   & 29.41 & 9.68  & 3.50 & 76.29  & 55.28 & 6.67 \\  
        \specialrule{0em}{1.pt}{0.5pt}
        \hline
        \specialrule{0em}{1.pt}{0.5pt}
           \textbf{\methodname} & 29.41 & \textbf{9.62}  & \textbf{3.50} & \textbf{79.41}  & \textbf{56.78} & \textbf{6.58} \\ 
        \bottomrule
        \end{tabular}
        }
    \vspace{-0.1cm}
    \caption{Ablation study of visual conditioning on LRS3 test set. }
    \label{tab:visual}
    \vspace{-0.05cm}
\end{table}

\paragraph{Ablation on visual conditioning.} 
To further explore our visual conditioning, we conduct ablation studies on the LRS3 benchmark in Table~\ref{tab:visual}. We evaluate the impact of the GE2E loss $\mathcal{L}_{\text{id}}$ by removing it from the training objective. The results reveal a substantial decline in speaker similarity from 56.78\% to 34.10\%, while WER remain unaffected. This confirms that the GE2E loss is indispensable for identity preservation, effectively guiding the model to extract implicit vocal timbre from facial cues. Second, we assess the Poster2~\cite{posterv2:journals/corr/abs-2301-12149} encoder by replacing it with Poster~\cite{DBLP:conf/iccvw/ZhengM023}. This substitution leads to a noticeable drop in emotion accuracy from 79.41\% to 76.29\%, validating the superiority of Poster2 in capturing fine-grained affective information.

\section{Conclusion}
\label{sec:conclusion} 

We present \methodname, a Hierarchical Codec Diffusion Transformer that redefines how visual features and speech tokens are aligned in VTS generation. By leveraging the hierarchy of discrete speech tokens, \methodname enables precise synchronization of lip motion and identity at lower levels, while capturing expressive emotional and prosodic dynamics at higher levels. We also design a dual-scale AdaLN, which effectively captures global vocal style and local prosody dynamics. 
Extensive experiments conducted on benchmark datasets, including LRS2 and LRS3, demonstrate the superiority of \methodname over state-of-the-art methods in terms of naturalness, expressiveness, and synchronization fidelity, establishing \methodname as a promising solution for real-world VTS applications.

{
    \small
    \bibliographystyle{ieeenat_fullname}

}

\end{document}